\documentclass[preprint,preprintnumbers, prd, floatfix, superscriptaddress,nofootinbib] {revtex4-1}
\usepackage{epsfig}
\usepackage{subfigure}
\usepackage{dcolumn}
\usepackage{bm}
\usepackage[usenames ,dvipsnames]{xcolor}
\usepackage{slashed}
\usepackage{graphicx,color}

\begin{document}
\title{Semileptonic baryonic $B$ decays}

\author{Yu-Kuo Hsiao}
\email{Email address: yukuohsiao@gmail.com}
\affiliation{School of Physics and Information Engineering, Shanxi Normal University, 
Taiyuan 030031, China}
\date{\today}

\begin{abstract}
We study the semileptonic $B\to{\bf B\bar B'}L\bar L'$ decays with
$\bf B\bar B'$ ($L\bar L'$) representing a baryon (lepton) pair.
Using the new determination of the $B\to{\bf B\bar B'}$ transition form factors,
we obtain ${\cal B}(B^-\to p\bar p \mu^-\bar \nu_\mu)
=(5.4\pm 2.0)\times 10^{-6}$ agreeing with the current data.
Besides, ${\cal B}(B^-\to \Lambda\bar p \nu \bar \nu)=(3.5\pm 1.0)\times 10^{-8}$ 
is calculated to be 20 times smaller than the previous prediction. In particular, we predict 
${\cal B}(\bar B^0_s\to 
p\bar \Lambda e^- \bar \nu_e,p\bar \Lambda \mu^- \bar \nu_\mu,p\bar \Lambda \tau^- \bar \nu_\tau)
=(2.1\pm 0.6,2.1\pm 0.6,1.7\pm 1.0)\times 10^{-6}$ and
${\cal B}(\bar B^0_s\to \Lambda\bar \Lambda \nu\bar \nu)=(0.8\pm 0.2)\times 10^{-8}$,
which can be accessible to the LHCb experiment.
\end{abstract}

\maketitle
\section{introduction}
In the non-leptonic baryonic $B$ decays,
the observation of 
$B\to p\bar p (\pi,K^{(*)},D^{(*)})$ and $B^-\to \Lambda\bar p (J/\psi,\gamma)$
suggests the unique existence of the $B\to{\bf B\bar B'}$ transition~\cite{Chua:2002wn,Geng:2006wz,Geng:2004jk},
with which the $CP$ asymmetries of $B^-\to p\bar p (\pi^-,K^{(*)-})$~\cite{Geng:2006jt,Geng:2007cw}
and the branching fractions of $B^-\to \Lambda\bar p D^{(*)0}$,
$\bar B^0\to \Sigma^0\bar \Lambda D^0$~\cite{Chen:2008sw}
have been predicted, and verified by the later measurements~\cite{pdg}.

The semileptonic $B$ decays of
$B^-\to p\bar p \ell^-\bar \nu_\ell$ and $B^-\to \Lambda\bar p\nu_\ell\bar \nu_\ell$
with $\ell$ denoting $e$, $\mu$ or $\tau$ can provide another evidence for
the $B\to{\bf B\bar B'}$ transition~\cite{Geng:2011tr,Geng:2012qn}. 
Like the studies of the semileptonic 
$B^-\to\pi^+\pi^-\ell^-\bar \nu_\ell$ decays~\cite{Tsai:2021ota,Belle:2020xgu},
the full dibaryon invariant mass spectrum can be used to test
the possible co-existence of the resonant and non-resonant contributions.
Therefore, we have predicted 
${\cal B}(B^-\to p\bar p e^- \bar \nu_e)
=(1.04\pm 0.26\pm 0.12)\times10^{-4}$~\cite{Geng:2011tr} and
${\cal B}(B^-\to \Lambda\bar p \nu\bar \nu)
=(7.9\pm 1.9)\times10^{-7}$~\cite{Geng:2012qn}.
We have also predicted ${\cal R}_{e/\mu}\equiv$
${\cal B}(B^-\to p\bar p e^- \bar \nu_e)/{\cal B}(B^-\to p\bar p \mu^- \bar \nu_\mu)\simeq 1$~\cite{Geng:2011tr}.
By contrast, the pole model argument leads to
the evaluation of ${\cal B}(B\to {\bf B\bar B'}\ell\bar \nu_\ell)=10^{-5}-10^{-6}$~\cite{Hou:2000bz}. 

Experimentally, 
it has been measured that~\cite{ppenu_cleo,Belle:2013uqr,LHCb:2019cgl,BaBar:2019awu}
\begin{eqnarray}\label{data0}
&&
{\cal B}_{ex}(B^- \to p\bar p e^-\bar \nu_e)
=(5.8\pm 3.7\pm 3.6)\times 10^{-4}\;
(<1.2\times 10^{-3})\;[\text{Cleo}]\,,\nonumber\\
&&
{\cal B}_{ex}(B^- \to p\bar p e^-\bar \nu_e)=
(8.2^{+3.7}_{-3.2}\pm 0.6)\times 10^{-6}\;[\text{Belle}]\,,\nonumber\\
&&
{\cal B}_{ex}(B^- \to p\bar p \mu^-\bar \nu_\mu)=
(3.1^{+3.1}_{-2.4}\pm 0.7)\times 10^{-6}\;[\text{Belle}]\,,\nonumber\\
&&
{\cal B}_{ex}(B^- \to p\bar p \mu^-\bar \nu_\mu)=
(5.27^{+0.23}_{-0.24}\pm 0.21\pm 0.15)\times 10^{-6}\;[\text{LHCb}]\,,\nonumber\\
&&
{\cal B}_{ex}(B^- \to \Lambda\bar p \nu\bar \nu)=
(0.4\pm 1.1\pm 0.6)\times 10^{-5}\;(<3.0\times 10^{-5})\;[\text{Babar}]\,.
\end{eqnarray}
The threshold effect commonly observed in $B\to{\bf B\bar B'}M$
is also observed in $B^- \to p\bar p \mu^-\bar \nu_e$~\cite{LHCb:2019cgl},
which is drawn as a peak around the threshold area 
of $m_{\bf B\bar B'}\simeq m_{\bf B}+m_{\bf \bar B'}$ in the ${\bf B\bar B'}$ invariant mass spectrum.
There is no sign that the $B$ to ${\bf B\bar B'}$ transition receives a resonant contribution.
Nonetheless, it is clearly seen that ${\cal B}_{ex}(B^- \to p\bar p \mu^-\bar \nu_\mu)$
is 20 times smaller than the prediction~\cite{Geng:2011tr}.
This has been pointed out as the theoretical challenge 
to alleviate the discrepancy~\cite{Huang:2021qld}. On the other hand,
the ratio ${\cal R}_{e/\mu}\simeq 1$ as a test of the lepton universality is not conclusive,
and the prediction of ${\cal B}(B^- \to \Lambda\bar p \nu\bar \nu)$ 
is within the experimental upper bound.

In Ref.~\cite{Chen:2008sw},
the $B\to{\bf B\bar B'}$ transition form factors ($F_{\bf B\bar B'}$)
are extracted with the data from $B\to{\bf B\bar B'}M$,
which cause the overestimation of ${\cal B}(B\to p\bar p\ell\bar \nu)$. 
With the same theoretical inputs, ${\cal B}(B^- \to \Lambda\bar p \nu\bar \nu)$ 
might be overestimated as well~\cite{Geng:2012qn}.
A question is hence raised: whether there exist 
the universal $B\to{\bf B\bar B'}$ transition form factors 
to explain the nonleptonic and semileptonic baryonic $B$ decays.

In this paper, we propose to perform a new global fit, 
in order to accommodate the current data of 
$B\to{\bf B\bar B'}L\bar L'$ with $L\bar L'$ denoting a lepton pair and $B\to{\bf B\bar B'}M$.
With $F_{\bf B\bar B'}$ determined from the new global fit,
we will re-investigate $B^- \to \Lambda\bar p \nu\bar \nu$.
Since LHCb has been able to accumulate more events for the $\bar B_s^0$ decays,
we will study $\bar B_s^0\to p\bar \Lambda \ell^- \bar \nu$ and
$\bar B_s^0\to\Lambda\bar \Lambda \nu\bar \nu$ decays for future measurements.

\section{Formalism}
%
\begin{figure}[t!]
\centering
\includegraphics[width=2.8in]{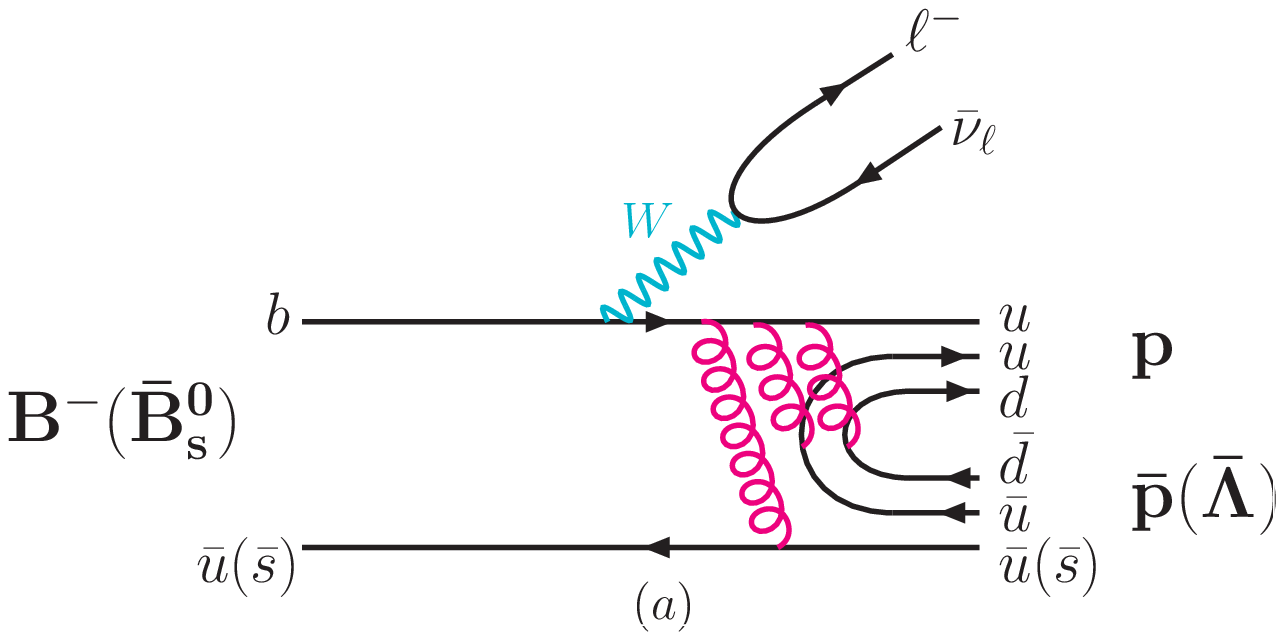}\\
\includegraphics[width=2.7in]{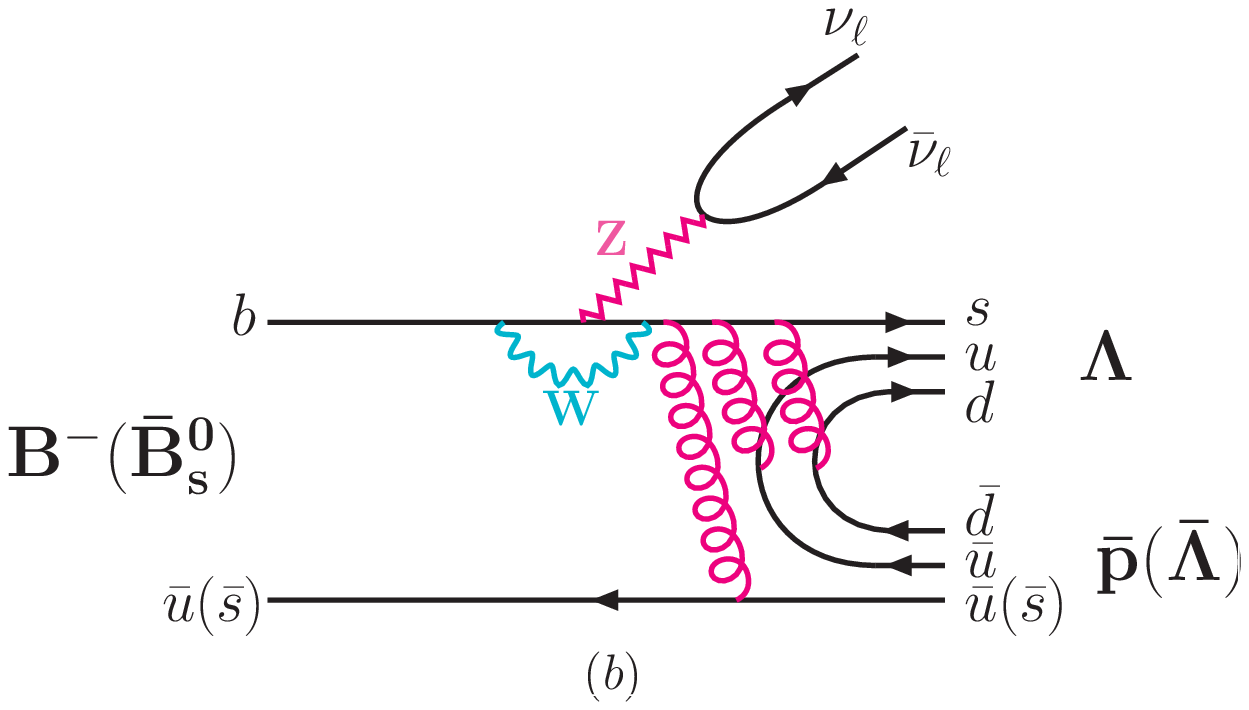}
\includegraphics[width=3.1in]{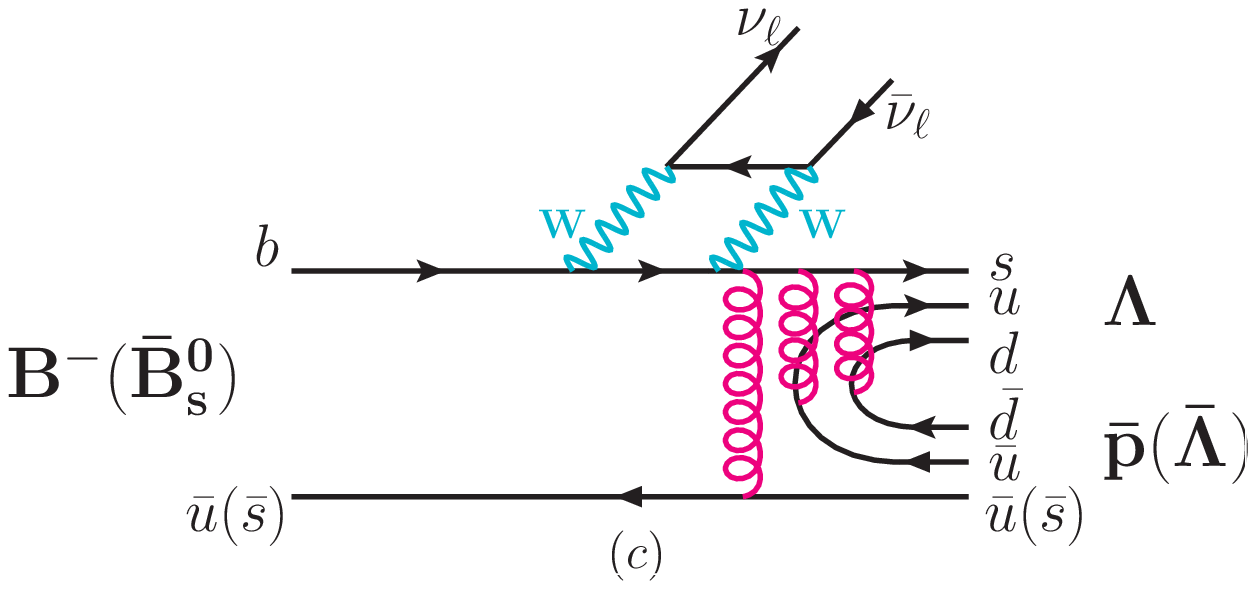}
\caption{Feynman diagrams for the $B\to {\bf B\bar B'}L\bar L'$ decays,
where $(a)$ depicts $B^-\to p\bar p\ell^-\bar \nu_\ell$ and $\bar B^0_s\to p\bar\Lambda\ell^-\bar \nu_\ell$, while
$(b,c)$ $B^-\to\Lambda\bar p\nu_\ell\bar \nu_\ell$ and $\bar B^0_s\to\Lambda\bar \Lambda\nu_\ell\bar\nu_\ell$.}\label{fig1}
\end{figure}
%
The semileptonic baryonic $B$ decays come from
the quark-level $b\to u\ell\bar\nu_\ell$ and $b\to s\nu_\ell \bar \nu_\ell$ processes.
In Fig.~\ref{fig1}a, $b\to u\ell\bar\nu_\ell$ appear 
as the tree-level $b\to uW,W\to\ell\bar\nu_\ell$ decays.
Due to the loop contributions from
the penguin-level $b\to sZ,Z\to\nu_\ell\bar \nu_\ell$ and box diagrams
in Figs.~\ref{fig1}b and~\ref{fig1}c, respectively~\cite{Hou:1986ug},
$b\to s\nu_\ell \bar \nu_\ell$ can be rarer than $b\to u\ell\bar\nu_\ell$.
The effective Hamiltonians for the above semileptonic $b$ decays
are given by~\cite{btosnunu,Hou:1986ug}
\begin{eqnarray}\label{effH}
&&
{\cal H}(b\to u \ell\bar \nu_\ell)=
\frac{G_F V_{ub}}{\sqrt 2}\;\bar u\gamma_\mu (1-\gamma_5)b\; \bar \ell\gamma^\mu (1-\gamma_5)\nu_\ell\,,\nonumber\\
&&
{\cal H}(b\to s \nu_\ell \bar \nu_\ell)=
\frac{G_F}{\sqrt 2}\frac{\alpha_{em}}{2\pi \text{sin}^2\theta_W }
\lambda_t D(x_t) \bar s\gamma_\mu(1-\gamma_5)b\bar \nu_\ell\gamma^\mu(1-\gamma_5)\nu_\ell\,,
\end{eqnarray}
where $G_F$ is the Fermi constant, $V_{ub}$ and $\lambda_t\equiv V_{ts}^* V_{tb}$
are the Cabibbo–Kobayashi– Maskawa (CKM) matrix elements, 
and $D(x_t)$ with $x_t\equiv m_t^2/m_W^2$ is the top-quark loop function~\cite{btosnunu,Hou:1986ug,Belanger:1990ur,Buchalla:1993bv}.
According to ${\cal H}(b\to u \ell\bar \nu_\ell,s \nu_\ell \bar \nu_\ell)$,
the amplitudes of $B\to{\bf B\bar B'}L\bar L'$ with $L\bar L'=(\ell\bar \nu_\ell,\nu_\ell\bar \nu_\ell)$
can be derived as~\cite{Geng:2011tr,Geng:2012qn}
\begin{eqnarray}\label{amp1}
&&{\cal M}(B\to {\bf B\bar B'} \ell^- \bar \nu_\ell)=\frac{G_F V_{ub}}{\sqrt 2}
\langle {\bf B\bar B'}|\bar u\gamma_\mu (1-\gamma_5)b|B
\rangle \;\bar \ell\gamma^\mu (1-\gamma_5) \nu_\ell\;,\nonumber\\
&&{\cal M}(B\to {\bf B\bar B'} \nu_\ell\bar \nu_\ell)=
\frac{G_F}{\sqrt 2}\frac{\alpha_{\text{em}}}{2\pi \text{sin}^2\theta_W }\lambda_t D(x_t)
\langle {\bf B\bar B'}|\bar s\gamma_\mu (1-\gamma_5)b|B\rangle
\;\bar \nu_\ell\gamma^\mu (1-\gamma_5) \nu_\ell\;,
\end{eqnarray}
with $\langle {\bf B\bar B'}|(\bar qb)|B\rangle$ representing
the matrix elements of the $B$ meson to ${\bf B\bar B'}$ transition. 
In Figs.~\ref{fig1}(a,b,c), $B\to{\bf B\bar B'}L\bar L'$ occur as
$B^-\to p\bar p\ell\bar \nu_\ell,\Lambda\bar p\nu_\ell\bar \nu_\ell$ and
$\bar B_s^0\to p\bar \Lambda\ell\bar \nu_\ell,\Lambda\bar \Lambda\nu_\ell\bar \nu_\ell$ for our study.

%
\begin{figure}[t!]
\centering
\includegraphics[width=2.6in]{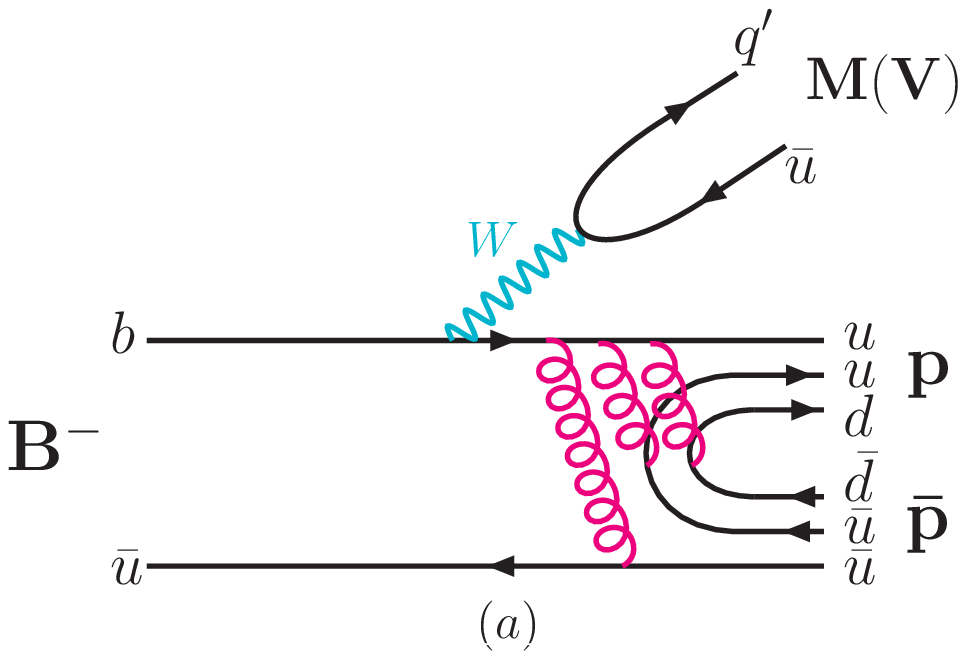}
\includegraphics[width=3.0in]{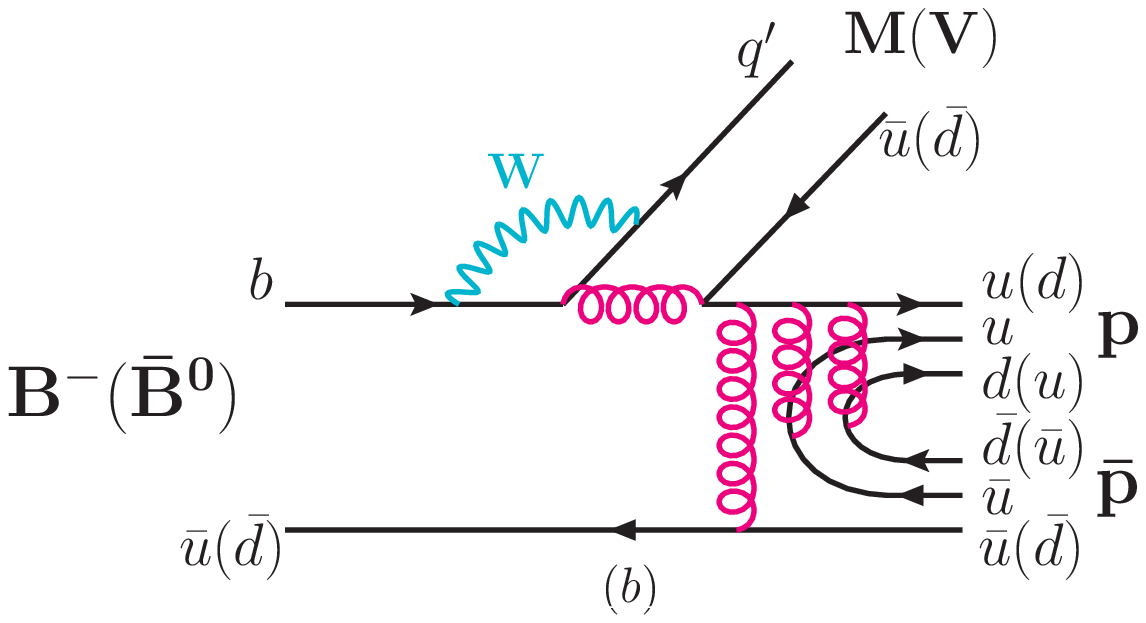}\\
\includegraphics[width=2.4in]{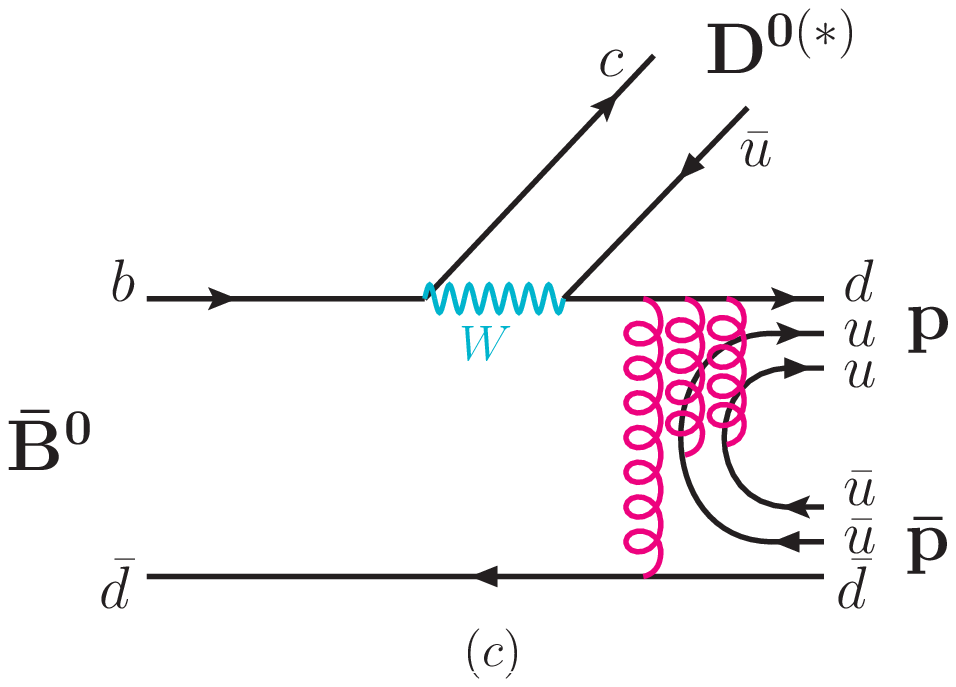}
\caption{Feynman diagrams for the three-body baryonic $B$ decays,
where $(a,b)$ and $(c)$ depict $B\to p\bar p M(V)$ and $\bar B^0\to p\bar p D^{0(*)}$,
respectively.}\label{fig1b}
\end{figure}
%
The amplitudes of the non-leptonic $B\to{\bf B\bar B'}M$ decays 
have two forms~\cite{Chua:2002wn,Chua:2001vh,Chua:2001xn}:
${\cal M}_C\propto \langle {\bf B\bar B'}|(\bar q q')|0\rangle$ $\times\langle M|(\bar q b)|B\rangle$
and ${\cal M}_T\propto \langle M|(\bar q q')|0\rangle\langle {\bf B\bar B'}|(\bar q b)|B\rangle$, where
${\cal M}_C$ 
denotes the current amplitude with $\bf B\bar B'$ produced from the quark current~\cite{Chua:2001vh,Chua:2001xn,Chua:2002yd,Huang:2022oli,Geng:2006yk}, and 
${\cal M}_T$ the transition amplitude 
with $\bf B\bar B'$ from the $B$ meson transition~\cite{Chua:2002wn,Geng:2006wz,Geng:2004jk}.
Clearly, ${\cal M}_T(B\to{\bf B\bar B'}M)$ and $B\to{\bf B\bar B'}L\bar L'$ can be related
by $\langle {\bf B\bar B'}|(\bar q b)|B\rangle$~\cite{Geng:2011tr,Geng:2012qn}. 
As seen in Fig.~\ref{fig1b},
$B\to p\bar p M$ with $M=(\pi,K)$, $B\to p\bar p V$ with $V=(\rho,K^{*})$, and
$\bar B^0\to p\bar p D^{0(*)}$ involve the transition amplitudes,
given by~\cite{Chua:2002wn,Hsiao:2016amt,Geng:2016fdw,Hsiao:2018umx}
\begin{eqnarray}\label{amp2}
&&
{\cal M}(B\to p\bar p M)=\frac{G_F}{\sqrt 2}(\hat {\cal M}_1+\hat {\cal M}_6)\,,\nonumber\\
&&
\hat {\cal M}_1=\alpha_1^{qq'}\langle M|\bar q'\gamma_\mu(1-\gamma_5)u|0\rangle 
\langle p\bar p|\bar q\gamma^\mu(1-\gamma_5)b|B\rangle\,,\nonumber\\
&&
\hat {\cal M}_6=\alpha_6^{qq'}\langle M|\bar q'(1+\gamma_5)u|0\rangle 
\langle p\bar p|\bar q(1-\gamma_5)b|B\rangle\,,
\nonumber\\
&&
{\cal M}(B\to p\bar p V)=\frac{G_F}{\sqrt 2}\alpha_1^{qq'}\langle V|\bar q'\gamma_\mu(1-\gamma_5)u|0\rangle 
\langle p\bar p|\bar q\gamma^\mu(1-\gamma_5)b|B\rangle\,,
\nonumber\\
&&{\cal M}(\bar B^0\to p\bar p D^{0(*)})=\frac{G_F}{\sqrt 2}V_{cb}V_{ud}^*a_2
\langle D^{0(*)}|\bar c\gamma_\mu(1-\gamma_5) u|0\rangle
\langle p\bar p|\bar d\gamma^\mu(1-\gamma_5) b|\bar B^0\rangle\,,
\end{eqnarray}
with $(q,q')=(u,d)$ for $B^-\to p\bar p \pi^-$ and $B^-\to p\bar p\rho^-$, 
$(q,q')=(u,s)$ for $B^-\to p\bar p K^{-}$ and $B^-\to p\bar p K^{*-}$, and
$(q,q')=(d,s)$ for $\bar B^0\to p\bar p \bar K^0$ and $\bar B^0\to p\bar p \bar K^{*0}$.
The parameters in Eq.~(\ref{amp2}) result from the factorization approach~\cite{ali},
written as
\begin{eqnarray}
&&\alpha_1^{uq'}=V_{ub}V_{uq'}^* a_1-V_{tb}V_{tq'}^*a_4\,,\nonumber\\ 
&&\alpha_1^{ds}=-V_{tb}V_{ts}^*a_4\,,\nonumber\\
&&\alpha_6^{uq'}=\alpha_6^{dq'}=V_{tb}V_{tq'}^*2a_6\,,
\end{eqnarray}
with $a_i=c^{eff}_i+c^{eff}_{i\pm 1}/N_c$ for $i=$odd (even),
where $c_i^{eff}$ are the effective Wilson coefficients, and $N_c$ the color number~\cite{ali}.

The matrix elements of the $B\to{\bf B\bar B'}$ transition in Eqs.~(\ref{amp1}) and (\ref{amp2})
can be presented as~\cite{Chua:2002wn,Geng:2006wz,Chen:2008sw,
Hsiao:2016amt,Geng:2016fdw,Huang:2021qld,Hsiao:2018umx}
\begin{eqnarray}\label{FFactor2}
\langle {\bf B\bar B'}|V^b_\mu|B\rangle&=&
i\bar u[  g_1\gamma_{\mu}+g_2i\sigma_{\mu\nu}p^\nu +g_3 p_{\mu}
+g_4(p_{\bf\bar B'}+p_{\bf B})_\mu +g_5(p_{\bf\bar B'}-p_{\bf B})_\mu]\gamma_5v\,,\nonumber\\
\langle {\bf B\bar B'}|A^b_\mu|B\rangle&=&
i\bar u[ f_1\gamma_{\mu}+f_2i\sigma_{\mu\nu}p^\nu +f_3 p_{\mu}
+f_4(p_{\bf\bar B'}+p_{\bf B})_\mu +f_5(p_{\bf\bar B'}-p_{\bf B})_\mu]v\,,\nonumber\\
\langle {\bf B\bar B'}|S^b|B\rangle&=&
i\bar u[ \bar g_1\slashed {p}+\bar g_2(E_{\bf \bar B'}+E_{\bf B})
+\bar g_3(E_{\bf \bar B'}-E_{\bf B})]\gamma_5v\,,\nonumber\\
\langle {\bf B}{\bf\bar B'}|P^b|B\rangle&=&
i\bar u[ \bar f_1\slashed {p}+\bar f_2(E_{\bf \bar B'}+E_{\bf B})
+\bar f_3(E_{\bf \bar B'}-E_{\bf B})]v\,,
\end{eqnarray}
with $V^b_\mu (A^b_\mu)\equiv \bar q\gamma_\mu(\gamma_5) b$, $S^b(P^b)\equiv \bar q b$,
and $p_\mu\equiv(p_B-p_{\bf B}-p_{\bf\bar B'})_\mu$. $F_{\bf B\bar B'}\equiv (g_i,f_i,\bar g_j,\bar f_j)$
with $i=1,2, ...,5$ and $j=1,2, 3$ are the $B\to{\bf B\bar B'}$ transition form factors.

$F_{\bf B\bar B'}$ are momentum dependent. 
In terms of perturbative QCD (pQCD) counting rules~\cite{Brodsky1,Brodsky2,Brodsky3,
Chua:2002wn,Geng:2006wz,Chen:2008sw,Hsiao:2016amt}, 
one parameterizes that
\begin{eqnarray}\label{figi}
&&
f_i=\frac{D_{f_i}}{t^3}\,,\;g_i=\frac{D_{g_i}}{t^3}\,,\;\nonumber\\
&&
\bar f_j=\frac{D_{\bar f_j}}{t^3}\,,\;\bar g_j=\frac{D_{\bar g_j}}{t^3}\,,
\end{eqnarray}
with $t\equiv (p_{\bf B}+p_{\bf\bar B'})^2$. For $F_{\bf B\bar B'}\propto1/t^n$,
$n=3$  corresponds to the three gluon propagators, which are drawn 
in Figs.~\ref{fig1}$(a,b,c)$ and Figs.~\ref{fig1b}$(a,b,c)$. Since $V_\mu^b$ and $A_\mu^b$ 
can be combined as the right-handed chiral current $R_\mu=(V_\mu^b+A_\mu^b)/2$, 
and the baryon decomposed of the right and left-handed states, that is,
$|{\bf B}_{R+L}\rangle=|{\bf B}_R\rangle+|{\bf B}_L\rangle$,
it leads to~\cite{Chen:2008sw,Huang:2021qld}
\begin{eqnarray}\label{J1J2}
&&
\langle {\bf B}_{R+L}{\bf\bar B'}_{R+L}|R_\mu|B\rangle=\nonumber\\
&&
i m_b\bar u\gamma_\mu\bigg[\frac{1+\gamma_5}{2}G_R +\frac{1-\gamma_5}{2}G_L\bigg]v+
i\bar u\gamma_\mu\not{\!p}_b\bigg[\frac{1+\gamma_5}{2}G^k_R +\frac{1-\gamma_5}{2}G^k_L\bigg]v\,,
\end{eqnarray}
where $|B_q\rangle\sim \bar b\gamma_5 q|0\rangle$ has been used.
As the chiral charge, $Q\equiv R_{\mu=0}$ annihilates the $b$ quark, and
creates a valence quark in $\bf B$, while the spectator quark in the $B$ meson
is transformed as a valence quark ($\bar q_i$) in $\bf \bar B'$. 
We hence obtain
$G_{R,L}^{(k)}$ as the $B\to{\bf B\bar B'}$ transition form factors in the chiral representation.  
When the chirality states of a spinor $(R,L)$
are taken as the helicity states $(\uparrow,\downarrow)$, 
one can see $\bar q_i$ with the helicity to be (anti-)parallel $[||(\overline{||})]$ to the helicity of ${\bf \bar B'}$,
such that the chiral charge acting on $\bar q_i$ 
can be more explicitly defined as $Q_{||(\overline{||})}(i)$ ($i=1,2,3$). 
We thus derive that
\begin{eqnarray}
G_{R(L)}&\propto&
e_{||}^{R(L)} G_{||}+e_{\overline{||}}^{R(L)} G_{\overline{||}}\,,
\nonumber\\
G^k_{R(L)}&\propto&
\bar e_{||}^{R(L)}G^k_{||}+\bar e_{\overline{||}}^{R(L)}G^k_{\overline{||}}\,,
\end{eqnarray}
where $e_{||}^{R(L)}$ and $e_{\overline{||}}^{R(L)}$ 
sum over the weight factors of ${\bf B}_{R(L)}{\bf\bar B'}_{R(L)}$,
and $\bar e_{||}^{R(L)}$ and $\bar e_{\overline{||}}^{R(L)}$ those of ${\bf B}_{L(R)}{\bf\bar B'}_{R(L)}$.
By defining $G_{||(\overline{||})}^{(k)}\equiv D_{||(\overline{||})}^{(k)}/t^3$ $(k=2,3, ...,5)$,
we relate the two sorts of the form factors as~\cite{Chua:2002wn,Geng:2006wz,Hsiao:2016amt}
\begin{eqnarray}\label{D1}
&&
D_{g_1}=\frac{5}{3}D_{||}-\frac{1}{3}D_{\overline{||}}\,,\;
D_{f_1}=\frac{5}{3}D_{||}+\frac{1}{3}D_{\overline{||}}\,,\;
D_{g_k}=\frac{4}{3}D_{||}^k=-D_{f_k}\,,\nonumber\\
&&
D_{g_1}=\frac{1}{3}D_{||}-\frac{2}{3}D_{\overline{||}}\,,\;
D_{f_1}=\frac{1}{3}D_{||}+\frac{2}{3}D_{\overline{||}}\,,\;
D_{g_k}=\frac{-1}{3}D_{||}^k=-D_{f_k}\,,\nonumber\\
&&
D_{g_1}=D_{f_1}=-\sqrt\frac{3}{2}D_{||}\,,\;
D_{g_k}=-D_{f_k}=-\sqrt\frac{3}{2}D_{||}^k\,,\nonumber\\
&&
D_{g_1}=D_{f_1}=\sqrt\frac{3}{2}D_{||}\,,\;
D_{g_k}=-D_{f_k}=\sqrt\frac{3}{2}D_{||}^{k}\,,\nonumber\\
&&
D_{g_1}=D_{f_1}=D_{||}\,,\;D_{g_k}=-D_{f_k}=-D_{||}^k\,,
\end{eqnarray}
for 
$\langle p\bar p|(\bar u b)|B^-\rangle$, 
$\langle p\bar p|(\bar d b)|\bar B^0\rangle$, 
$\langle \Lambda\bar p|(\bar s b)|B^-\rangle$,
$\langle p\bar \Lambda|(\bar u b)|\bar B^0_s\rangle$, and
$\langle \Lambda\bar \Lambda|(\bar s b)|\bar B^0_s\rangle$,
respectively. Likewise,
we perform a derivation for $\bar g_j$ ($\bar f_j$)
through the (pseudo-)scalar current,
which leads to~\cite{Hsiao:2016amt,Hsiao:2018umx}
\begin{eqnarray}\label{D2}
&&
D_{\bar g_1}=\frac{5}{3}\bar D_{||}-\frac{1}{3}\bar D_{\overline{||}}\,,\;
D_{\bar f_1}=\frac{5}{3}\bar D_{||}+\frac{1}{3}\bar D_{\overline{||}}\,,\;
D_{\bar g_{2,3}}=\frac{4}{3}\bar D_{||}^{2,3}=-D_{\bar f_{2,3}}\,,\nonumber\\
&&
D_{\bar g_1}=\frac{1}{3}\bar D_{||}-\frac{2}{3}\bar D_{\overline{||}}\,,\;
D_{\bar f_1}=\frac{1}{3}\bar D_{||}+\frac{2}{3}\bar D_{\overline{||}}\,,\;
D_{\bar g_{2,3}}=\frac{-1}{3}\bar D_{||}^{2,3}=-D_{\bar f_{2,3}}\,,
\end{eqnarray}
for 
$\langle p\bar p|(\bar u b)|B^-\rangle$ and
$\langle p\bar p|(\bar d b)|\bar B^0\rangle$, respectively.
Note that $R(L)\sim \uparrow(\downarrow)$ is based on the approximation with 
the large energy transfer, which is conveniently presented as $t\to\infty$.
It is also derived that the correction term is of order $m_q/\sqrt t$~\cite{Brodsky1,Brodsky2,Brodsky3}.
In fact, $\sqrt t$ of a few GeV has been large enough to suppress the correction term~\cite{Brodsky3}.
Consequently, the relations with the chirality (helicity) symmetry are shown 
to be able to describe the scattering processes~\cite{Brodsky3}. 
For the baryonic $B$ decays, $\sqrt t >2$~GeV is also sufficient 
for the holding of the relations in Eqs.~(\ref{D1}) and (\ref{D2}).

\begin{figure}[t!]
\centering
\includegraphics[width=3.6in]{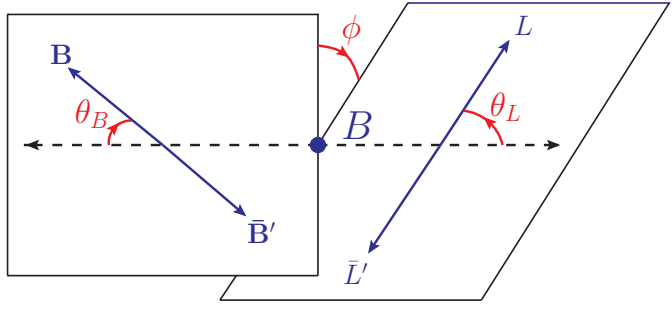}
\caption{The angular variables $\theta_{\bf B}$, $\theta_L$ and $\phi$
depicted for the four-body $B\to{\bf B\bar B'}L\bar L'$ decays.}\label{fig2}
\end{figure}
%
The four-body 
$B(p_B)\to {\bf B}(p_{\bf B}) {\bf\bar B'}(p_{\bf\bar B'})L(p_L)\bar L' (p_{\bar L'})$ decay
involves five kinematic variables in the phase space, that is,
$s\equiv (p_{L}+p_{\bar L'})^2\equiv m_{L\bar L'}^2$, $t$, 
and $(\theta_{\bf B},\theta_{\bf L},\phi)$~\cite{Kl4,Wise,Hsiao:2017nga}. 
As depicted in Fig.~\ref{fig2},
the angle $\theta_{\bf B(L)}$ is between $\vec{p}_{\bf B}$  ($\vec{p}_{L}$)
in the $\bf B\bar B'$ ($L\bar L'$) rest frame and 
the line of flight of the $\bf B\bar B'$ ($L\bar L'$) system in the $B$ meson rest frame.
The angle $\phi$ is from the $\bf B\bar B'$ plane to the $L\bar L'$ plane
defined by the momenta of the $\bf B\bar B'$ pair and $L\bar L'$ pair
in the $B$ meson rest frame, respectively.
The partial decay width then reads~\cite{Geng:2011tr,Geng:2012qn}
\begin{eqnarray}\label{Gamma1}
d\Gamma=\frac{|\bar {\cal M}|^2}{4(4\pi)^6 m_B^3}X
\alpha_{\bf B}\alpha_{\bf L}\, ds\, dt\, d\text{cos}\,\theta_{\bf B}\, d\text{cos}\,\theta_{\bf L}\, d\phi\,,
\end{eqnarray}
where 
$X=[(m_B^2-s-t)^2/4-st]^{1/2}$,
$\alpha_{\bf B}=\lambda^{1/2}(t,m_{\bf B}^2,m_{\bf \bar B'}^2)/t$, and
$\alpha_{\bf L}=\lambda^{1/2}(s,m_{L}^2,m_{\bar L'}^2)/s$,
with $\lambda(a,b,c)=a^2+b^2+c^2-2ab-2bc-2ca$. For integration,
the allowed ranges of the five variables are
$(m_L+m_{\bar L'})^2\leq s\leq (m_{B}-\sqrt{t})^2$,
$(m_{\bf B}+m_{\bf \bar B'})^2\leq t\leq (m_{B}-m_L-m_{\bar L'})^2$,
$0\leq \theta_{\bf B,L}\leq \pi$, and $0\leq \phi\leq 2\pi$.
The partial decay width of $B(p_B)\to{\bf B}(p_{\bf B}){\bf \bar B'}(p_{\bf\bar B'})M(p_M)$
involves two variables in the phase space, given by~\cite{Geng:2006wz,Hsiao:2016amt}
\begin{eqnarray}\label{Gamma2}
d\Gamma=
\frac{\beta_{\bf B}^{1/2}\beta^{1/2}_t}{(8\pi m_B)^3}
|\bar {\cal M}|^2\;dt\;d\text{cos}\theta\,,
\end{eqnarray}
where $\beta_{\bf B}=[1-(m_{\bf B}+m_{\bf \bar B'})^2/t][1-(m_{\bf B}-m_{\bf\bar B'})^2/t]$,
$\beta_t=[(m_B+m_M)^2-t][(m_B-m_M)^2-t]$, and
$\theta$ is the angle between the meson and baryon moving directions in the $\bf B\bar B'$ rest frame.  
The allowed regions 
of the variables are $-1<\cos\theta<1$ and $(m_{\bf B}+m_{\bf \bar B'})^2<t<(m_B-m_M)^2$.
For the global fit in the next section, 
we define the $CP$ asymmetry~\cite{Hsiao:2019ann,Geng:2006jt},
and angular asymmetries of $B\to{\bf B\bar B'}M$~\cite{Geng:2006wz,Hsiao:2016amt,Huang:2022oli}
and $B\to{\bf B\bar B'}L\bar L'$~\cite{Geng:2011tr,Geng:2012qn},
written as
\begin{eqnarray}
{\cal A}_{CP}&\equiv& \frac
{\Gamma(B\to {\bf B\bar B'} M)-\Gamma(\bar B\to {\bf B\bar B'} \bar M)}
{\Gamma(B\to {\bf B\bar B'} M)+\Gamma(\bar B\to {\bf B\bar B'} \bar M)}\,,\nonumber\\
{\cal A}_{FB,\theta_i}&\equiv& \frac{\Gamma(\cos\theta_i>0)-\Gamma(\cos\theta_i<0)}
{\Gamma(\cos\theta_i>0)+\Gamma(\cos\theta_i<0)}\,,
\end{eqnarray}
where $\bar B\to {\bf B\bar B'} \bar M$ represents the anti-particle decay.

\section{Numerical Results and Discussions}
In the numerical analysis, the CKM matrix elements in the Wolfenstein parameterization 
read~\cite{pdg}
\begin{eqnarray}
&&
V_{ub}=A\lambda^3(\rho-i\eta)\,,
V_{ud}=1-\lambda^2/2\,,
V_{us}=\lambda,\,\nonumber\\
&&
V_{cb}=A\lambda^2\,,
V_{tb}=1\,,
V_{td}=A\lambda^3\,,
V_{ts}=-A\lambda^2\,,
\end{eqnarray}
with $(\lambda,A,\rho,\eta)=(0.225,0.826,0.163\pm 0.010,0.357\pm 0.010)$.
%
\begin{table}[b]
\caption{The effective Wilson coefficients $c_i^{eff}$ ($i=1,2,\,...,6$) for $b$ and $\bar b$ decays.}\label{effWC}
{\tiny
\begin{tabular}{|c|cc|}
\hline
$c_i^{eff}$ &$b\to d\;(\bar b\to\bar d)$&$b\to s\;(\bar b\to\bar s)$\\
\hline\hline
$c_1^{eff}$ 
& $1.168\,(1.168)$ 
&  $1.168\,(1.168)$\\
$c_2^{eff}$   
& $-0.365\,(-0.365)$ 
&  $-0.365\,(-0.365)$\\
$10^4 c_3^{eff}$ 
& $238.0 + 12.7i\,(257.4 + 46.1i)$ 
& $243.3 + 31.2i\,(240.9 + 32.3i)$\\
$10^4 c_4^{eff}$   
& $-497.0 - 38.0i\,(-555.2 - 138.3i)$ 
& $-512.8 - 93.7i\,(-505.7 - 96.8i)$\\
$10^4 c_5^{eff}$  
& $145.5 + 12.7i\,(164.7 + 46.1i)$ 
& $150.7 + 31.2i\,(148.4 + 32.3i)$\\
$10^4 c_6^{eff}$ 
& $-633.8 - 38.0i\,(-692.0 - 138.3i)$ 
& $-649.6 - 93.7i\,(-642.6 - 96.8i)$\\
\hline
\end{tabular}}
\end{table}
%
From Refs.~\cite{btosnunu,Hou:1986ug,Belanger:1990ur,Buchalla:1993bv}, 
we adopt $D(x)$ as
\begin{eqnarray}
D(x)&=&D_0(x)+\frac{\alpha_s}{4\pi}D_1(x)\,,\nonumber\\
D_0(x)&=&\frac{x}{8}\bigg[-\frac{2+x}{1-x}+\frac{3x-6}{(1-x)^2}ln(x)\bigg]\,,\nonumber\\
D_1(x)&=&-\frac{23x+5x^2-4x^3}{3(1-x)^2}
+\frac{x-11x^2+x^3 +x^4}{(1-x)^3}ln(x)
+\frac{8x+4x^2 +x^3-x^4}{2(1-x)^3}ln^2(x)\nonumber\\
&&-\frac{4x-x^3}{(1-x)^2}L_2(1-x)
+8x\frac{\partial D_0(x)}{\partial x}ln(\mu^2/m_W^2)\,,
\end{eqnarray}
where $L_2(1-x)\equiv \int^x_1  ln(t)/(1-t) dt$ and $\mu=m_b$. 
For $B\to p\bar p M(V)$ and $\bar B^0\to p\bar p D^{0(*)}$, we present $c_i^{eff}$ in Table~\ref{effWC},
where $b$ and $\bar b$ decays are both considered,
together with the decay constants
$(f_\pi,f_K,f_\rho,f_{K^*})=
(130.2\pm 1.2,155.7\pm 0.3,210.6\pm 0.4,204.7\pm 6.1)$~MeV~\cite{Hsiao:2014mua,pdg}
and $(f_D,f_{D^*})=(208.9\pm 6.5,252.2\pm 22.7)$~MeV~\cite{Lucha:2014xla,Hsiao:2016amt}.
In the generalized edition of the factorization~\cite{ali,Hsiao:2019ann}, $N_c$ is taken as
the effective color number with $N_c^{(eff)}=(2,3,\infty)$,
in order that the non-factorizable QCD corrections can be estimated. 
%
\begin{table}[b]
\caption{Experimental data for the $B^-\to p\bar p\ell^-\nu_\ell$ and
$B\to p\bar p M_{(c)}$ decays, where the notation~$\dagger$ for ${\cal A}_{FB}$ 
denotes the contribution from $m_{p\bar p}<2.85$~GeV, and
${\cal B}(B^-\to p\bar p \mu^- \bar \nu_\mu)$ has combined the Belle and LHCb data
in Eq.~(\ref{data0}).}\label{data1}
{
\tiny
\begin{tabular}{|l|cl|}
\hline
$\;\;\;\;\;\;\;\;$ Decay modes &Data&\\\hline
\hline
$10^6 {\cal B}(B^-\to p\bar p e^- \bar \nu_e)$
&$8.2\pm 3.8$&\cite{Belle:2013uqr}\\
$10^6 {\cal B}(B^-\to p\bar p \mu^- \bar \nu_\mu)$
&$5.2\pm 0.4$&\cite{Belle:2013uqr,LHCb:2019cgl}\\
\hline
$10^6 {\cal B}(B^-\to p\bar p \pi^-)$
&$1.62\pm 0.20$&\cite{pdg}\\
$10^6 {\cal B}(B^-\to p\bar p K^-)$
&$5.9\pm 0.5$&\cite{pdg}\\
$10^ 6{\cal B}(\bar B^0\to p\bar p \bar K^0)$
&$2.66\pm 0.32$&\cite{pdg}\\
$10^2 {\cal A}_{CP}(B^-\to p\bar p \pi^-)$
&$0\pm 4$&\cite{pdg}\\
$10^2 {\cal A}_{CP}(B^-\to p\bar p K^-)$
&$0\pm 4$&\cite{pdg}\\
$10^2 {\cal A}_{FB}(B^-\to p\bar p \pi^-)$
&$(-40.9\pm 3.4)^\dagger$&\cite{LHCb:2014nix}\\
$10^2 {\cal A}_{FB}(B^-\to p\bar p K^-)$
&$(49.5\pm 1.4)^\dagger$&\cite{LHCb:2014nix}\\
\hline
$10^6 {\cal B}(B^-\to p\bar p \rho^-)$
&$4.6\pm 1.3$&\cite{pdg}\\
$10^6 {\cal B}(B^-\to p\bar p K^{*-})$
&$3.4\pm 0.8$&\cite{Belle:2008zkc}\\
$10^6 {\cal B}(\bar B^0\to p\bar p \bar K^{*0})$
&$1.2\pm 0.3$&\cite{Belle:2008zkc}\\
$10^2 {\cal A}_{CP}(B^-\to p\bar p K^{*-})$
&$21\pm 16$&\cite{pdg}\\
\hline
$10^4 {\cal B}(\bar B^0\to p\bar p D^0)$
&$1.04\pm 0.07$&\cite{pdg}\\
$10^4 {\cal B}(\bar B^0\to p\bar p D^{*0})$
&$0.99\pm 0.11$&\cite{pdg}\\
\hline
\end{tabular}}
\end{table}
%
Using the minimum $\chi^2$-fit of
\begin{eqnarray}\label{fitEQ}
\chi^2&=&
\sum\bigg(\frac{{\cal O}_{th}^i-{\cal O}_{ex}^i}{\sigma_{ex}^i}\bigg)^2+
\bigg(\frac{{|V_{ub}|}_{th}-{|V_{ub}|}_{ex}}{\sigma_{{|V_{ub}|}_{ex}}}\bigg)^2\,,
\end{eqnarray}
we test if the observables of non-leptonic and semileptonic baryonic $B$ decays
can both be interpreted, where ${\cal O}_{th}^i$ stand for the theoretical calculations of
${\cal B}$, ${\cal A}_{CP}$ and ${\cal A}_{FB}$, while
${\cal O}_{ex}^i$ the experimental inputs in Table~\ref{data1}, 
together with $\sigma_{ex}^i$ the experimental errors.
Since the $V_{ub}$ in Eq.~(\ref{amp1}) is for  the exclusive baryonic $B_{(s)}$ decays,
which can be different from that in the inclusive ones~\cite{Hsiao:2018zqd,Hsiao:2017umx,Hsiao:2015mca},
we choose $|V_{ub}|_{ex}=(3.43\pm 0.32)\times 10^{-3}$ 
determined from the $\bar B_s^0$ and baryonic $\Lambda_b$ decays~\cite{pdg}
as our experimental input in Eq.~(\ref{fitEQ}).

With 16 experimental inputs from Table~\ref{data1} and $|V_{ub}|_{ex}$,
we fit $(D_{||},D_{\overline{||}},D_{2,3,4,5})$ and 
$(\bar D_{||},\bar D_{\overline{||}},\bar D_{2,3})$ in Eqs.~(\ref{D1}) and (\ref{D2}), 
respectively, and $|V_{ub}|_{th}$, which amount to 11 parameters, such that 
the number of degrees of freedom denoted by $d.n.f$ is counted as $d.n.f=16-11=5$.
As a result, we obtain $\chi^2/n.d.f=1.86$ as a measure of the global fit, 
and extract that\\
\begin{eqnarray}\label{extraction}
&&
(D_{||},D_{\overline{||}})=(11.2\pm 43.5,332.3\pm 17.2)\;{\rm GeV}^{5}\,,\nonumber\\
&&
(D_{||}^2,D_{||}^3,D_{||}^4,D_{||}^5)
=(47.7\pm 10.1,442.2\pm 103.4,-38.7\pm 9.6, 80.7\pm 27.2)\;{\rm GeV}^{4}\,,\nonumber\\
&&
(\bar D_{||},\bar D_{\overline{||}},\bar D_{||}^2,\bar D_{||}^3)
=(-59.9\pm 12.9,23.8\pm 6.8, 90.9\pm 11.1, 131.7\pm 330.7)\;{\rm GeV}^{4}\,,
\end{eqnarray}
with $N_c^{eff}=2$ and $\infty$ for $B\to p\bar p M(V)$ and $B\to p\bar p D^{0(*)}$, respectively. 
Using the parameters in Eq.~(\ref{extraction}), 
we calculate the branching fractions and angular asymmetries of 
$B^-\to p\bar p\ell\bar \nu,\Lambda\bar p\nu\bar \nu$ and
$\bar B_s^0\to p\bar \Lambda\ell\bar \nu,\Lambda\bar \Lambda\nu\bar \nu$,
of which the results are compared with the experimental data in Table~\ref{fit_result}.
We also draw the ${p\bar p}$ invariant mass spectrum
for $B^-\to p\bar p\mu^-\bar \nu_\mu$ in Fig.~\ref{fig:spectrum}.
%
\begin{table}[b!]
\caption{Our calculations for the semileptonic $B\to{\bf B\bar B'}L\bar L'$ decays. 
For ${\cal B}(B\to{\bf B\bar B'}\ell\bar\nu_\ell)$,
the values in the parentheses correspond to $\ell=(e,\mu,\tau)$, where
the first and second errors come from $|V_{ub}|$ and the form factors in Eq.~(\ref{extraction}), 
respectively. For ${\cal B}(B\to{\bf B\bar B'}\nu\bar \nu)=\Sigma_\ell{\cal B}(B\to{\bf B\bar B'}\nu_\ell\bar \nu_\ell)$
and ${\cal A}_{FB}(B\to{\bf B\bar B'}L\bar L')$, 
the errors take into account the uncertainties of the form factors in Eq.~(\ref{extraction}).}\label{fit_result}
{
\tiny
\begin{tabular}{|l|c|c|}
\hline
$\;\;\;\;\;\;\;\;\;\;$ Decay modes &This work & Data\\\hline
\hline
$10^6 {\cal B}(B^-\to p\bar p \ell^- \bar \nu_\ell)$
&$(5.3\pm 1.1\pm 1.7,5.4\pm 1.1\pm 1.7,7.6\pm 1.5\pm 3.9)$
&$(8.2\pm 3.8$~\cite{Belle:2013uqr},
$5.2\pm 0.4$~\cite{Belle:2013uqr,LHCb:2019cgl},---$)$\\
$10^2{\cal A}_{FB,\theta_{\bf B}}(B^-\to p\bar p \ell^- \bar \nu_\ell)$  
&$(1.4\pm 12.6,1.4\pm 12.6,1.4\pm 12.6)$ &--- \\
$10^2{\cal A}_{FB,\theta_{\bf L}}(B^-\to p\bar p \ell^- \bar \nu_\ell)$   
&$(-41.7\pm 21.4,-41.2\pm 20.4,-2.1\pm 5.0)$&--- \\
\hline
$10^6 {\cal B}(\bar B^0_s\to p\bar \Lambda \ell^- \bar \nu_\ell)$
&$(2.1\pm 0.4 \pm 0.5,2.1\pm 0.4 \pm 0.5,1.7\pm 0.3\pm 0.9)$
&---\\
$10^2{\cal A}_{FB,\theta_{\bf B}}(\bar B^0_s\to p\bar \Lambda \ell^- \bar \nu_\ell)$  
&$(25.7\pm 11.4,25.0\pm 11.3,-3.5\pm 2.7)$ &--- \\
$10^2{\cal A}_{FB,\theta_{\bf L}}(\bar B^0_s\to p\bar \Lambda \ell^- \bar \nu_\ell)$   
&$(-44.1\pm 10.8,-43.7\pm 10.3,0.4\pm 5.5)$&--- \\
\hline
\hline
${\cal B}(B^-\to \Lambda\bar p \nu\bar \nu)$
&$(3.5\pm 1.0)\times 10^{-8}$
&$(0.4\pm 1.3)\times 10^{-5}\;(<3\times 10^{-5})$~\cite{BaBar:2019awu}\\
$10^2{\cal A}_{FB,\theta_{\bf B}}(B^-\to \Lambda\bar p \nu\bar \nu)$  
&$22.8\pm 11.2$ 
&--- \\
$10^2{\cal A}_{FB,\theta_{\bf L}}(B^-\to \Lambda\bar p \nu\bar \nu)$   
&$-40.9\pm 8.3$
&--- \\
\hline
${\cal B}(\bar B^0_s\to \Lambda\bar \Lambda \nu\bar \nu)$
&$(0.8\pm 0.2)\times 10^{-8}$
&---\\
$10^2 {\cal A}_{FB,\theta_{\bf B}}(\bar B^0_s\to \Lambda\bar \Lambda \nu\bar \nu)$  
&$24.4\pm 11.8$ 
&--- \\
$10^2 {\cal A}_{FB,\theta_{\bf L}}(\bar B^0_s\to \Lambda\bar \Lambda \nu\bar \nu)$   
&$-40.1\pm 8.0$ 
&--- \\
\hline
\end{tabular}}
\end{table}
%
\section{Discussions and Conclusions}
%
\begin{figure}[t!]
\centering
\includegraphics[width=2.85 in]{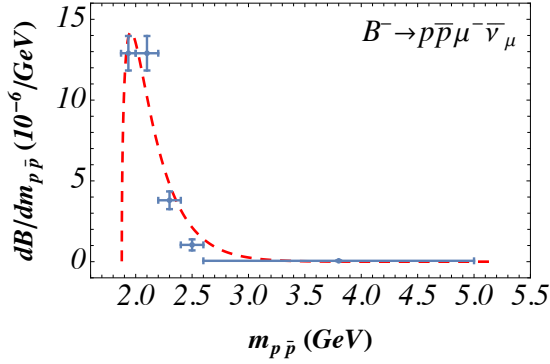}
\caption{The $p\bar p$ invariant mass spectrum of $B^-\to p\bar p\mu^-\bar \nu_\mu$, 
where the data points are from LHCb~\cite{LHCb:2019cgl}.}\label{fig:spectrum}
\end{figure}
%
Since $\chi^2/n.d.f=1.86$ presents a reasonable fit, it indicates that
the most recent data in Table~\ref{data1} can be explained. 
It is interesting to note that
${\cal B}(B^-\to p\bar p \pi^-,p\bar p\rho^-)$~\cite{Geng:2006wz,Geng:2006jt}
were once overestimated~\cite{pdg,LHCb:2014nix,Belle:2007oni}, and 
the relation of ${\cal A}_{FB}(B^-\to p\bar p\pi^-)\simeq {\cal A}_{FB}(B^-\to p\bar p K^-)$~\cite{Geng:2006wz}
was not verified by the measurements~\cite{LHCb:2014nix,Belle:2007oni}.
This is due to $F_{\bf B\bar B'}$ determined by the $B\to p\bar p K$ data~\cite{Geng:2006wz},
while $B\to p\bar p K$ are in fact the penguin dominated decays
with $\hat {\cal M}_6\propto\langle p\bar p|(S-P)^b|B\rangle$ to give the main contribution.
To avoid the inconsistency unable to be solved at that time, 
one performed the extraction of Ref.~\cite{Chen:2008sw} that excluded 
${\cal B}(B^-\to p\bar p K^-)$, ${\cal B}(\bar B^0\to p\bar p \bar K^0)$, 
and ${\cal A}_{FB}(B^-\to p\bar p K^-)$, in order that 
the more associated tree dominated decays of
$B\to p\bar p(\pi,\rho)$, $\bar B^0\to p\bar p D^{0(*)}$, and $B\to{\bf B\bar B}L\bar L'$
can be studied. However, it resulted in an unsatisfactory global fit not to accommodate the all data.

As $F_{\bf B\bar B'}$ determined in this work can be universal 
for the non-leptonic and semileptonic decay channels, we calculate
${\cal B}(B^-\to p\bar p e^-\bar \nu_e)=(5.3\pm 2.0)\times 10^{-6}$ and
${\cal B}(B^-\to p\bar p\mu^-\bar \nu_\mu)=(5.4\pm 2.0)\times 10^{-6}$
agreeing with the experimental values. Moreover,
we revisit $B^-\to\Lambda\bar p\nu\bar \nu$, and obtain
${\cal B}(B^-\to\Lambda\bar p\nu\bar \nu)=(3.5\pm 1.0)\times 10^{-8}$
20 times smaller than the number of Ref.~\cite{Geng:2012qn}.

Like the theoretical illustration in $B\to{\bf B\bar B'}$ 
and $B\to{\bf B\bar B'}M$~\cite{Suzuki:2006nn,Hsiao:2018umx}, 
the gluon propagators of Figs.~\ref{fig1}(a,b,c) play the key role
in the ${\bf B\bar B'}$ formation of $B\to{\bf B\bar B'}L\bar L'$, 
where two of them provide the valence quarks in $\bf B\bar B'$,
while the another one speeds up the spectator quark in $B$. Accordingly, 
the approach of pQCD counting rules derives that $F_{\bf B\bar B'}\propto 1/t^3$.

One can test the momentum dependence  of $B^-\to p\bar p\mu^-\bar \nu_\mu$,
which is by scanning the partial branching fraction
as a function of $\sqrt t=m_{p\bar p}$. In Fig.~\ref{fig:spectrum},
as we draw the line to agree with the five data points~\cite{LHCb:2019cgl}; 
particularly, those around the area of $\sqrt t\sim m_{\bf B}+m_{\bf\bar B'}$
for the threshold effect, it is shown that $F_{\bf B\bar B'}$ as a function of $1/t$ 
can describe $B\to{\bf B\bar B'}L\bar L'$.

By normalizing the prediction of the pQCD model~\cite{Geng:2011tr}, 
LHCb draws the $m_{p\bar p}$ spectrum of $B^-\to p\bar p\mu\bar\nu$ in Fig.~4 of Ref.~\cite{LHCb:2019cgl},
where the line is higher and narrower than our result. 
The difference is caused by the fact that the line of of Ref.~\cite{LHCb:2019cgl} 
is chosen to more agree with the two data points around $m_{p\bar p}\sim 2.5$~GeV.
Subsequently, the peak should reach $17\times 10^{-6}$ 
to be above the data point around $m_{p\bar p}\sim 2$~GeV
for integrating over the partial branching fraction as large as ${\cal B}\simeq 5\times 10^{-6}$.
In comparison, our result prefers to agree with the threshold data points; however,
requiring some broadening to give a sufficient branching fraction.

The decay channel $\bar B^0_s\to \Lambda\bar p K^+(\bar \Lambda p K^-)$ is
the first observation of a baryonic $\bar B_s^0$ decay~\cite{LHCb:2017mfl}, 
whose branching fraction 
${\cal B}(\bar B^0_s\to \Lambda\bar p K^+ +\bar \Lambda p K^-)=5.46\times 10^{-6}$
is as large as those of the three-body baryonic $B^-$ ($\bar B^0$) decays.
Hence, the semileptonic baryonic $\bar B^0_s$ decay is supposed to be
compatible with $B^-\to p\bar p\ell\bar \nu_\ell$. In our prediction, 
we present 
\begin{eqnarray}
&&{\cal B}(\bar B^0_s\to p\bar \Lambda e^- \bar \nu_e,p\bar \Lambda \mu^- \bar \nu_\mu)
=(2.1\pm 0.6,2.1\pm 0.6)\times 10^{-6}\,,
\end{eqnarray}
which are accessible to the LHCb experiment, whereas
${\cal B}(\bar B^0_s\to \Lambda\bar \Lambda \nu\bar \nu)=(0.8\pm 0.2)\times 10^{-8}$
is relatively small.

Because of $m_\tau\gg m_{e,\mu}$ that strongly shrinks the phase space,
it is anticipated that
${\cal B}(B\to{\bf B\bar B'}\tau\bar\nu)\ll{\cal B}(B\to{\bf B\bar B'}e\bar\nu,{\bf B\bar B'}\mu\bar\nu)$. 
Nonetheless, the amplitude of Eq.~(\ref{amp1}) and the matrix elements of Eq.~(\ref{FFactor2}) 
result in
\begin{eqnarray}
&&
i\bar u(g_3\gamma_5-f_3)v\; m_\ell\bar u_\ell\gamma_\mu (1+\gamma_5) v_{\bar\nu}\,
\end{eqnarray}
in ${\cal M}(B\to {\bf B\bar B'} \ell\bar\nu)$, where
$m_\tau$ is able to enhance the decay. We thus obtain
\begin{eqnarray}
&&{\cal B}(B^-\to p\bar p \tau^- \bar \nu_\tau)
=(7.6\pm 4.2)\times 10^{-6}\,,\;\nonumber\\
&&{\cal B}(\bar B^0_s\to p\bar \Lambda \tau^- \bar \nu_\tau)
=(1.7\pm 1.0)\times 10^{-6}\,,
\end{eqnarray}
as large as their counterparts. Likewise, the mass effect can be found in
${\cal M}(B\to\tau\bar \nu)\propto 
m_\tau \bar u_\tau(1+\gamma_5)v_{\bar \nu}$~\cite{Bevan:2014iga,Hou:2019uxa} and 
${\cal M}(B\to{\bf B}_c\bar {\bf B}')\propto 
m_c \langle {\bf B}_c\bar {\bf B}'|\bar c(1+\gamma_5)q|0\rangle$~\cite{Hsiao:2019wyd}, 
where $m_\tau$ and $m_c$ alleviate the decays from helicity suppression.

We study the angular asymmetries of the semileptonic $B\to{\bf B\bar B'}L\bar L'$ decays.
While ${\cal A}_{FB,\theta_{\bf B}}(B^-\to p\bar p \ell^- \bar \nu_\ell)$ are around several percents, 
${\cal A}_{FB,\theta_{\bf B}}(\bar B^0_s\to p\bar \Lambda e^- \bar \nu_e, p\bar \Lambda \mu^- \bar \nu_\mu)$ and 
${\cal A}_{FB,\theta_{\bf B}}(\bar B^0_s\to \Lambda\bar \Lambda \nu\bar \nu)$
can be around 25\%. 
Like the three-body baryonic $B$ decays~\cite{Geng:2006wz,Hsiao:2016amt,Huang:2022oli},
this implies a theoretical sensitivity for $F_{\bf B\bar B'}$ to be confirmed by future measurements.

In summary, 
we have investigated the semiletonic $B^-(\bar B_s^0)\to{\bf B\bar B'}L\bar L'$ decays
with $L\bar L'=(\ell\bar \nu_\ell,\nu\bar \nu)$. We have newly extracted 
the $B\to{\bf B\bar B'}$ transition form factors with the global fit that includes the data of 
$B\to p\bar p M(V)$, $\bar B^0\to p\bar p D^{0(*)}$ and $B\to p\bar p e^-\bar \nu_e,p\bar p\mu^-\bar \nu_\mu$ decays. 
In our demonstration,
${\cal B}(B^-\to p\bar p e^-\bar \nu_e,p\bar p\mu^-\bar \nu_\mu)$ once overestimated 
to be as large as $10^{-4}$ has been reduced to be around $5\times 10^{-6}$,
in agreement with the current data. We have also presented 
${\cal B}(B^-\to\Lambda\bar p\nu\bar \nu)=(3.5\pm 1.0)\times 10^{-8}$.
It has been found that 
${\cal B}(\bar B^0_s\to p\bar \Lambda e^- \bar \nu_e,p\bar \Lambda \mu^- \bar \nu_\mu,
p\bar \Lambda \tau^- \bar \nu_\tau)
=(2.1\pm 0.6,2.1\pm 0.6,1.7\pm 1.0)\times 10^{-6}$ and 
${\cal B}(\bar B^0_s\to \Lambda\bar \Lambda \nu\bar \nu)=(0.8\pm 0.2)\times 10^{-8}$
can be promising for future measurements.

\section*{ACKNOWLEDGMENTS}
The author would like to thank Prof.~C.~K.~Chua for useful discussions.
This work was supported by NSFC (Grants No.~11675030 and No.~12175128).


\end{document}